# Spin-dependent Forward Particle Correlations in p+p Collisions at $\sqrt{s}$ = 200 GeV

Nikola Poljak[a] for the STAR collaboration

[a]University of Zagreb, Croatia

**Abstract.** The STAR collaboration has reported precision measurements of the transverse single spin asymmetries for the production of the $\pi^0$ mesons from polarized proton collisions at $\sqrt{s}$ = 200 GeV. These measurements were obtained using modular forward detectors. The Forward Meson Spectrometer (FMS), covering a region of 2.5 < η < 4.0, and its engineering prototype, provide increased acceptance, as needed for spin-dependent correlation studies that could disentangle contributions to the forward $\pi^0$ asymmetries. We report on the status of the analysis.

**Keywords:** p p, colliding beams, polarized beam, hadroproduction, small-angle, STAR, single spin asymmetry, transverse momentum, $x_F$-dependence, electromagnetic calorimeter, 200 GeV

**PACS:** 13.88.+e; 13.85.Ni; 12.38.Qk

## INTRODUCTION

For a particle produced in a collision of transversely polarized protons, the analyzing power ($A_N$) is defined as the difference of spin-up and spin-down cross sections divided by their sum. $A_N$ is just one of many examples of transverse single spin asymmetries. $A_N$ is expected to be near zero in a leading-twist and collinear perturbative QCD (pQCD) description of particle production. Cross sections for pions produced with large Feynman-x ($2p_L/\sqrt{s}$) and moderate $p_T$ in p+p collisions are found to be in agreement with next-to-leading order pQCD calculations at $\sqrt{s}$ =200 GeV [1,2]. Precision measurements of the asymmetry as a function of $x_F$ and $p_T$ were reported, showing large $A_N$ at large $x_F$ [3]. Also, it was shown that the $x_F$ dependence matches the Sivers effect [4,5] expectations qualitatively. The $p_T$ dependence at fixed $x_F$ is not consistent with expectations of p-QCD based calculations.

Since these reports, there was a significant number of developments both in theory and experiment. New phenomenological analyses within a generalized parton model can now explain Sivers moments in semi-inclusive deep inelastic scattering and many features of p↑+p → π +X [6]. Previous expectations that the Collins effect [7] is suppressed in p↑+p → π +X were found incorrect due to a sign error [8]. COMPASS analyzed 20 percent of their transversely polarized proton data and reported non-zero Collins moments and Sivers moments compatible with zero, although the expected Sivers moments are small in the $(x,Q^2)$ range of their experiment. To conclude, the need still remains to separate Collins and Sivers effects in p↑+p → π +X.

The Sivers effect manifests itself as an asymmetry in the forward jet or gamma production while the Collins as an asymmetry in the forward jet fragmentation. To distinguish between the two one has to look beyond inclusive π events. Since the Collins effect is an azimuthal modulation of hadrons around the thrust axis of an outgoing quark, integrating over the azimuthal angle cancels it, leaving only the Sivers effect. This is one possible path to distinguishing the two effects.

## EXPERIMENTAL SETUP

The detectors used for the measurements done at STAR in the previous runs measured the inclusive π cross section as well as the single beam spin asymmetry for their inclusive production. These detectors, called Forward Pion Detector (FPD), were modular electromagnetic detectors, covering regions of $\langle\eta\rangle$ = 3.3, 3.7 or 4.0. The Forward Meson Spectrometer (FMS) now provides STAR with full coverage of the azimuth in the range $2.5 \leq \eta \leq 4.0$. The forward acceptance of STAR has been increased 20x by the FMS. Studies of inclusive production in p+p and d(p)+Au collisions over a broad range of $x_F$ and $p_T$ for mesons that decay electromagnetically to observed photons, electrons and positrons are now enabled. Also possible are particle correlation studies within a single forward jet and correlation studies of fragments from jet pairs.

The FMS is built from lead-glass detectors of two types: 788 cells of square cross section 5.8 cm on a side and 18 radiation lengths, and 476 cells of square cross section 3.8 cm on a side and 18 radiation lengths. The FMS spans a 2x2 $m^2$ area perpendicular to the beam and its front face is 706 cm from the interaction point.

The methodologies that were developed to calibrate the FPD have been adapted to calibrate the FMS. Individual detector conversion gains are iteratively adjusted to give the Gaussian centroid at the π mass in di-photon invariant mass distributions associated with each detector. Events used for calibration were required to satisfy a hardware-level individual detector threshold condition and have at least two photons reconstructed within a fiducial volume of the calorimeter. The two highest energy photons in an event are used to compute the mass. The energy sharing, defined as $|E_1 - E_2|/(E_1+E_2)$, must be < 0.7 for the small cells. In addition to adjustments of linear calibration factors for each detector, energy dependent corrections are found to be required. The centroid of the π peak in the mass distributions is found to increase with leading photon energy for both the data and full GEANT simulations. For the latter, it is found that when energy dependent corrections are applied to give the correct centroid for the pion peak in di-photon invariant mass distributions binned in leading photon energy, the reconstructed di-photon energy, on average, agrees with the generated pion energy.

The calibrations proved to be stable at the level of a few percent, as determined by analysis of the time dependence of the mass peak centroid. Another tool checking the calibration stability was a light-emitting diode (LED) monitoring system placed on the front face of the detector. This system had the capability of flashing a number of LED signals at a given time so a response from the detector could be read out.

Full PYTHIA/GEANT simulations have been made with adequate statistics to study neutral pion production to moderate $x_F$ at large $p_T$. Energy-binned mass distributions from data reconstructions are in reasonable agreement with similar distributions from reconstruction of simulation data.

# GOALS AND FIRST RESULTS

It is anticipated that the FMS will provide new information about transverse single spin asymmetries. To make a point of contact with previous work, a first analysis of the $x_F$ dependence of the $A_N$ for the $p\uparrow+p \rightarrow \pi +X$ from the FMS was performed.

To verify patterns of polarization signs for the colliding beam bunches, analysis of concurrently measured spin asymmetries from the STAR beam-beam counter (BBC) was performed. It was previously known that the azimuthal distribution of multiplicity in the BBC has a significant analyzing power [9], making it useful as a colliding beam polarimeter. Spin asymmetries from the BBC, averaged over all bunches, and bunch-by-bunch spin asymmetries from the BBC for each fill, confirm the veracity of the spin directions of the polarized colliding beams for each fill.

In the analysis of $p\uparrow+p \rightarrow \pi +X$, we divided the FMS into octants for inclusive $\pi$ spin sorting. A cross ratio analysis of the resulting distributions from opposite octants was then made. Analyzing powers comparable to prior measurements are found, from analysis of 75% of the 2008 run data, as shown in Fig. 1(b). Quantitative comparison of this data with earlier measurements requires binning in both $x_F$ and $p_T$. The dependence of the asymmetry on $<\cos \varphi>$, where $<\cos \varphi>$ is determined from data, is well represented by a linear function, as expected. The zero crossing in Fig. 1. (a) is shifted from $<\cos \varphi>$ at the level of $2\sigma$. Statistical errors are included in the result, with systematical errors still being evaluated. A first estimate based on significance distributions [3] yields $\sigma_{tot.} \leq 1.2\ \sigma_{stat.}$ The figures presented here are a different presentation of the same results shown at the SPIN-2008 symposium.

"Jet-like" events were also looked at. "Jet-like" clusters are formed in an event by considering energy depositions > 0.4 GeV in all cells of the FMS. A cluster consists of N cells, where N is the maximal subset of cells that are found to be within a cone of radius 0.5 in the $\eta$-$\varphi$ space. The $x_F$ and $p_T$ for the cluster are given by the vector sum of momenta from each cell, assuming the energy deposition is from photons originating from the collision vertex. Clusters with at least 15 cells with deposited energy > 0.4 GeV having cluster $p_T$ > 1 GeV/c and $x_F$ >0.2 are required in the analysis. A requirement that the cluster centroid is within the calorimeter volume by at least two cell widths is also imposed.

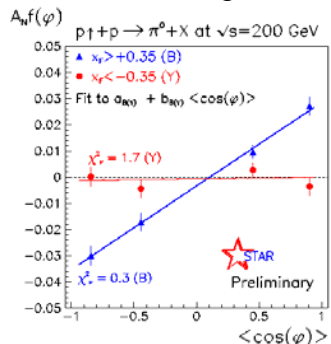
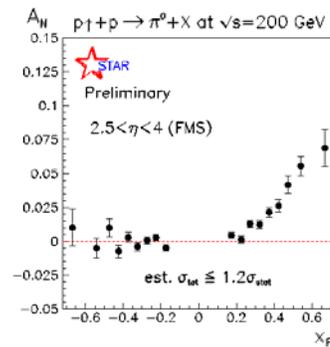

**FIGURE 1.** (a) $A_N$ versus $<\cos \varphi>$, as determined from data. The positive $x_F$ fit gives a $\varphi$-independent fit parameter that is non-zero at the level of $2\sigma$, possibly due to a statistical fluctuation.
(b) Variation of $A_N$ as a function of $x_F$ integrated over the FMS acceptance. The negative $x_F$ values are consistent with zero. The indicated uncertainties are statistical.

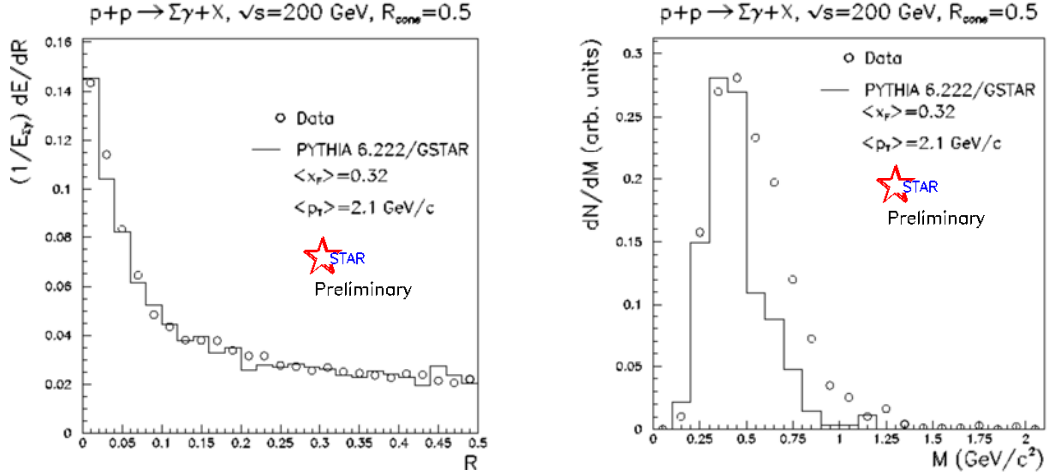

**FIGURE 2.** (a) "Jet-shape" distribution compared in data and simulation. The image shows the deposition of energy within the jet as a function of distance from the jet axis.

(b) Invariant mass distribution of "jet-like" clusters. Fair agreement is found between data and simulations, although the mass distribution is broader in data than from PYTHIA/GSTAR simulations.

The clustering analysis is applied to both data and PYTHIA/GEANT simulations. Good agreement for the jet-shape is found, and fair agreement for the jet-mass is found, as shown in Fig. 2.

## CONCLUSIONS AND OUTLOOK

The FMS is complete, in place, commissioned and operated in the 2008 run. It has 20 times more acceptance than the previous modular detectors. The reconstruction and calibration procedures were successfully ported from the FPD to the FMS. The calibration is mostly complete and shows good agreement with the simulated sample of events. The inclusive pion $A_N(x_F)$ is comparable to prior FPD precision measurements and the analysis of "jet-like" events is under way.

We aim to finish the analysis of "jet-like" events and extend the inclusive pion $A_N(p_T)$ measurements. Also, we want to determine $A_N$ for the final state that contains $\pi$ pairs as well as for states with heavier mesons. In the future runs, the goal is to go beyond $\pi$ detection to direct photons+jet final state $A_N$.

## REFERENCES

1. J. Adams et al., *Physical Review Letters* **97**, 152302 (2006)
2. I. Arsene et al., *Physical Review Letters* **98**, 252001 (2007)
3. B. I. Abelev et al., *Physical Review Letters* **101**, 222001 (2008)
4. D. Sivers, *Physical Review D* **41**, 83-90 (1990)
5. D. Sivers, *Physical Review D* **43**, 261-263 (1991)
6. M. Boglione, U. D'Alesio and F. Murgia, *Physical Review D* **77**, 051502 (2008)
7. J. C. Collins, *Nuclear Physics B* **396**, 161-182 (1993)
8. M. Anselmino et al., *Physical Review D* **71**, 014002 (2005)
9. J. Kiryluk et al., *hep-ex 0501072*, published in Spin 2004 Conference Proceedings, Trieste, Italy